\documentclass[3p,onecolumn]{elsarticle}

\usepackage{graphicx}
\usepackage{dcolumn}
\usepackage{pifont}
\usepackage{bm}
\usepackage{multirow}
\usepackage{amsmath}
\usepackage{float}
\usepackage{txfonts}
\usepackage[version=3]{mhchem}

\usepackage[dvips]{color}

\journal{Electrochimica Acta}

\begin{document}
\title{Critical role of water in defect aggregation and chemical degradation of perovskite solar cells}

\author[kimuniv-m]{Yun-Hyok Kye}
\author[kimuniv-m]{Chol-Jun Yu\corref{cor}}
\ead{ryongnam14@yahoo.com}
\author[kimuniv-m]{Un-Gi Jong}
\author[hk]{Yue Chen}
\author[icl]{Aron Walsh}

\cortext[cor]{Corresponding author}

\address[kimuniv-m]{Department of Computational Materials Design, Faculty of Materials Science, Kim Il Sung University, \\ Ryongnam-Dong, Taesong District, Pyongyang, Democratic People's Republic of Korea}
\address[hk]{Department of Mechanical Engineering, The University of Hong Kong, Pokfulam Road, Hong Kong SAR, China}
\address[icl]{Department of Materials, Imperial College London, London SW7 2AZ, United Kingdom}

\begin{abstract}
The chemical stability of methylammonium lead iodide (\ce{MAPbI3}) under humid conditions remains the primary challenge facing halide perovskite solar cells. We investigate defect processes in the water-intercalated iodide perovskite (\ce{MAPbI3}\_\ce{H2O}) and monohydrated phase (\ce{MAPbI3}$\cdot$\ce{H2O}) within a first-principles thermodynamic framework. We consider the formation energies of isolated and aggregated vacancy defects with different charge states under I-rich and I-poor conditions. It is found that a \ce{PbI2} (partial Schottky) vacancy complex can be formed readily, while the \ce{MAI} vacancy complex is difficult to form in the hydrous compounds. Vacancies in the hydrous phases create deep charge transition levels, indicating the degradation of halide perovskite upon exposure to moisture. Electronic structure analysis supports a novel mechanism of water-mediated vacancy-pair formation.
\end{abstract}

\maketitle

Low-cost perovskite solar cells (PSCs) based on methylammonium lead iodide (\ce{CH3NH3PbI3} or \ce{MAPbI3}) are rapidly evolving, with a record power-conversion efficiency (PCE) from under 4\% in 2009~\cite{Kojima} to over 22\% in recent years~\cite{WSYang}. However, PSCs have a critical problem of easy degradation by extrinsic as well as intrinsic factors, still preventing their outdoor installation~\cite{Wang,Berhe,Li}. In particular, the facile decomposition of \ce{MAPbI3} upon exposure to moisture has been recognized to be the major extrinsic factor of PSC degradation~\cite{Manser,WHuang,JHuang}. In fact, the PCE of \ce{MAPbI3} solar cells drops by nearly 90\% in a few days under an ambient environment ($T= 300$ K, relative humidity (RH) = 30$-$50\%)~\cite{You}, while \ce{MAPbI3} can be decomposed into MAI, \ce{PbI2}, and HI in a few hours at high humidity conditions~\cite{Niu}.

For a chemical explanation of this phenomenon, hydrolysis of \ce{MAPbI3} was initially suggested as the main mechanism, and based on the first-principles calculations, deprotonation of \ce{MA+} by \ce{H2O} was proposed as the principal cause of the hydrolysis~\cite{NiuRev,Frost14,Zhao}. Soon afterwards, however, it was demonstrated that \ce{MAPbI3} readily transformed to the monohydrate phase \ce{MAPbI3$\cdot$H2O} at moderate humidity (RH $\leq$ 60\%), while to the dihydrate phase \ce{(MA)4PbI6$\cdot2$H2O} at high humidity  (RH $\geq$ 80\%), at the initial stage of the \ce{MAPbI3} water-mediated decomposition process, which could be reversed by drying treatment~\cite{JYang,Christians,Leguy,Hao,Song16,Shirayama,NAhn,IJeon}. This can be explained by the hydrogen bonding interaction between the lead iodide framework and the organic \ce{MA+} cations in the perovskite crystal being weakened upon its hydration~\cite{Zhang15,Quarti,Gottesman}. \ce{MA+} can readily diffuse and separate from the \ce{PbI6} octahedra, resulting in a rapid decomposition of \ce{MAPbI3}. The activation barrier for vacancy-mediated \ce{MA+} migration was confirmed to be reduced from 1.18 eV in \ce{MAPbI3} to 0.38 eV in water-intercalated and 1.14 eV in monohydrated phases~\cite{Eames,Haruyama,Azpiroz,yucj16}. Although there have been some theoretical studies of the intrinsic point defects in \ce{MAPbX3} (X = I, Br, Cl)~\cite{Walsh,JKim,Buin15,Yin1,Du}, those in the hydrate phases remain unexplored.

In this Letter, we investigate the origin of perovskite decomposition through point defect processes in water-intercalated \ce{MAPbI3}, denoted as \ce{MAPbI3\_H2O} hereafter, and monohydrated phase, \ce{MAPbI3\cdot H2O}. The water-intercalated  \ce{MAPbI3\_H2O} is suggested as an intermediate phase during the transition to the hydrated phases due to the relatively low activation energies for water insertion into the perovskite surface (0.27 eV~\cite{Koocher} or 0.31 eV~\cite{Tong}), as well as for water molecular diffusion within the bulk crystal (0.28 eV~\cite{yucj16}). A density functional theory (DFT) approach combined with \textit{ab initio} thermodynamics is utilized to describe defect formation and interactions. Electrostatic stabilization by water is found to play a key role in defect clustering and ultimately in the stability of perovskites in humid environments. 

In the first stage, we performed structural optimizations of pristine \ce{MAPbI3},  water-intercalated \ce{MAPbI3}\_\ce{H2O}, and monohydrated \ce{MAPbI3$\cdot$H2O}. The lattice constant and bandgap of \ce{MAPbI3} were calculated to be 6.33 \AA~and 1.53 eV, which are in good agreement with the experimental values of 6.32$-$6.33 \AA~\cite{Poglitsch,Weller} and 1.50 eV~\cite{HSKim}. For the case of \ce{MAPbI3}\_\ce{H2O}, the unit cell containing a water molecule in the large interstitial space formed by a huge \ce{PbI6} framework became triclinic after optimization. The initial structure of \ce{MAPbI3$\cdot$H2O} with a monoclinic crystalline lattice and experimentally identified atomic positions~\cite{Imler} was also optimized, giving the lattice constants of $a=10.46$ \AA, $b=4.63$ \AA, $c=11.10$ \AA~and $\beta=101.50^\circ$ agreed well with the experimental values~\cite{Imler}. The bandgaps were calculated to be 1.86 eV in \ce{MAPbI3\_H2O}, and 2.47 eV in \ce{MAPbI3$\cdot$H2O}, being comparable with the previous DFT value of 2.52 eV for the monohydrated phase~\cite{Shirayama} and experimental value~\cite{Leguy}.

\begin{figure}[!t]
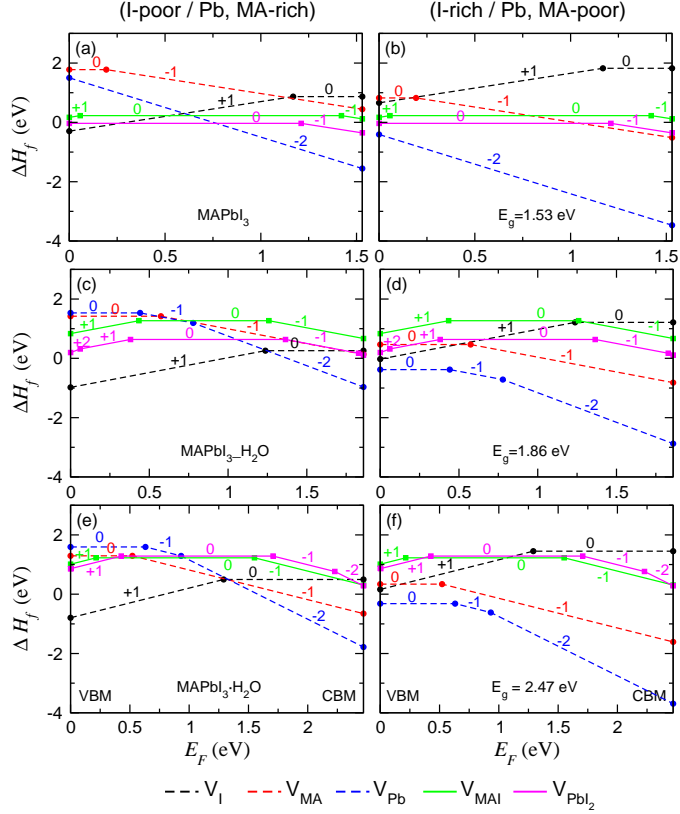

\begin{center}
\includegraphics[clip=true,scale=0.51]{fig1ab.eps}
\includegraphics[clip=true,scale=0.51]{fig1cd.eps}
\includegraphics[clip=true,scale=0.51]{fig1ef.eps}
\end{center}
\caption{\label{fig_dig}Formation enthalpies of vacancy point and pair defects as a function of the Fermi level ($E_F$) under I-poor (Pb, MA-rich) conditions (left panel) and I-rich (Pb, MA-poor) conditions (right panel) in (a)/(b) cubic \ce{MAPbI3}, (c)/(d) water-intercalated phase \ce{MAPbI3}\_\ce{H2O}, and (e)/(f) monohydrate phase \ce{MAPbI3$\cdot$H2O}. $E_F$ ranges within the bandgap ($E_g$) from the valence band maximum (VBM), set to 0 eV, to the conduction band minimum (CBM).}
\end{figure}

In the second stage, using the optimized unit cells, we built $(3\times3\times3)$ supercells for \ce{MAPbI3} (324 atoms) and \ce{MAPbI3\_H2O} (405 atoms), while the $(2\times3\times2)$ supercell for \ce{MAPbI3$\cdot$H2O} (360 atoms), with and without vacancy defects, and performed atomic relaxations with the fixed lattice constants (see Figures S1-S3). Isolated vacancy point ($V_I$, $V_{\ce{MA}}$, $V_{\ce{Pb}}$) and pair defects ($V_{\ce{MAI}}$, $V_{\ce{PbI2}}$) were created. 
For each vacancy defect, various charge states were considered to identify the thermodynamic charge transition levels. Figure~\ref{fig_dig} presents the defect formation energy diagrams at I-poor (Pb-rich and MA-rich) and I-rich (Pb-poor and MA-poor) conditions. These conditions correspond to an iodine precursor in orthorhombic solid form and a lead precursor in fcc solid form, respectively.
%

Amongst the vacancy point defects, lead vacancies with a charge state of $-2$ ($V^{-2}_{\ce{Pb}}$) in these three compounds, and additionally $-1$ charge state ($V^{-1}_{\ce{Pb}}$) and neutral state ($V^0_{\ce{Pb}}$) in the case of hydrous compounds, have the lowest formation energies in the whole range of Fermi level ($E_F$) at I-rich conditions. For I-poor growth, meanwhile, the iodine vacancies with a charge state of $+1$ ($V^{+1}_I$) have the lowest formation energies in the lower part of $E_F$, whereas the $V^{-2}_{\ce{Pb}}$ in the higher range of $E_F$. Note that for the case of \ce{MAPbI3} our results are consistent with the previous DFT results~\cite{Buin15,Yin1}, with some minor numerical differences due to the inclusion of dispersion corrections in this work.

Under I-poor conditions, MA vacancies with a neutral state ($V^0_{\ce{MA}}$) and a charge state of $-1$ ($V^{-1}_{\ce{MA}}$) have typically higher formation energies than $V_{\ce{Pb}}$ and $V_I$ in \ce{MAPbI3} and \ce{MAPbI3\_H2O}, but in-between values are found in the case of \ce{MAPbI3$\cdot$H2O}. 

\begin{table}[!b]
\caption{\label{tab_ene}Formation ($H_f$) and Binding ($E_b$) Energies of Schottky-type Vacancy Pair Defects in \ce{MAPbI3}, \ce{MAPbI3\_H2O} and \ce{MAPbI3$\cdot$H2O} (unit: eV per defect).}
\small
\begin{tabular}{lcc@{}ccc@{}ccc}
\hline 
 & \multicolumn{2}{c}{\ce{MAPbI3}} && \multicolumn{2}{c}{\ce{MAPbI3\_H2O}} && \multicolumn{2}{c}{\ce{MAPbI3$\cdot$H2O}} \\
\cline{2-3}\cline{5-6}\cline{8-9}
 & $H_f$ & $E_b$ && $H_f$ & $E_b$ && $H_f$ & $E_b$ \\
\hline
$V^0_{\ce{MAI}}$  & ~~~0.23 & 1.45~~ && 1.27 & $-$0.25~~ && 1.23 & $-$0.21 \\
$V^0_{\ce{PbI2}}$ & $-$0.03 & 0.94~~ && 0.64 & ~~~0.16~~ && 1.29 & ~~~0.28 \\
\hline
\end{tabular}
\end{table}

For vacancy pair defects, which can be viewed as compensated partial Schottky-type aggregates, we considered various charge states.  We calculated the binding energy defined as $E_b=H_f[\ce{A}]+H_f[\ce{B}]-H_f[\ce{AB}]$~\cite{Freysoldt}. 
Table~\ref{tab_ene} summarizes the formation and binding energies of the neutral  pairs of $V^0_{\ce{MAI}}$ and $V^0_{\ce{PbI2}}$ (for $E_b$ of charged pairs, see Table S1). For the case of \ce{MAPbI3}, the formation energy of $V^0_{\ce{MAI}}$ is 0.23 eV, which is much lower than 1.80 eV reported by Kim et al.~\cite{JKim}. If we use the MAI molecule instead of MAI solid as they did, it becomes 1.98 eV in better agreement. $V^0_{\ce{PbI2}}$ has the formation energy of $-0.03$ eV, being slightly lower than 0.03 eV reported by Kim et al.~\cite{JKim}, possibly due to a different crystal lattice. In general, the formation energies of these complex defects in the hydrous compounds are higher than those in the pristine \ce{MAPbI3}.

The formation of $V^0_{\ce{PbI2}}$ in all the compounds is more favorable than the formation of the individual vacancy point defects $V^{+1}_I$ and $V^{-2}_{\ce{Pb}}$ due to their positive binding energies (Table~\ref{tab_ene}). Therefore, it is expected that $V^{+1}_I$ and $V^{-2}_{\ce{Pb}}$ are formed first (they are dominant defects), and then the interaction between them leads to the formation of $V^0_{\ce{PbI2}}$ independently of whether the hydrous compound or not. Water adsorption into the perovskite crystal reduces the activation barrier for vacancy-mediated \ce{I-} ion migration~\cite{yucj16}, resulting in an enhancement of $V^0_{\ce{PbI2}}$ formation. As shown in Figure~\ref{fig_dig}, the formation energy of $V^{+1}_{\text{I}}$ ($V^{-2}_{\ce{Pb}}$) at the I-rich condition is higher (lower) than at the I-poor condition (their concentrations have a reverse feature), and thus the reaction speed of $V^0_{\ce{PbI2}}$ formation can be lower at the I-rich condition (\ce{Pb2+} ion migration is quite difficult). Experimentally, the I-rich condition can be realized by directly adding \ce{I_3^-} in solution~\cite{WSYang}. In indirect ways, increasing \ce{PbI2} concentration relative to MAI concentration as in many other experimental works~\cite{HSKo} can effectively inhibit the formation of $V^0_{\ce{PbI2}}$ from the decomposition of \ce{MAPbI3} from the viewpoint of chemical equilibrium of reaction, and furthermore, formation of a passivating \ce{MAPbI3}/\ce{PbI2} interface~\cite{Supasai,LWang} also provide \ce{PbI2} excess environment to prevent the forward reaction for $V^0_{\ce{PbI2}}$ formation. Surprisingly, the binding energies of $V^0_{\ce{MAI}}$ in the hydrous compounds are negative, although it is positive in the pristine perovskite. This indicates that in the hydrous compounds other products such as \ce{HI}, \ce{CH3NH2} and \ce{I2} rather than MAI can be formed during chemical decomposition.

%
\begin{figure}[!t]
\begin{center}
\includegraphics[clip=true,scale=0.52]{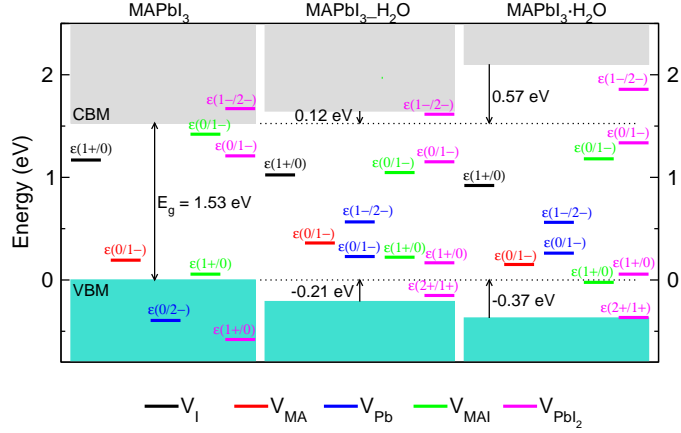}
\end{center}
\caption{\label{fig_trans}Band alignment and thermodynamic transition levels in  \ce{MAPbI3}, water-intercalated \ce{MAPbI3}\_\ce{H2O} and monohydrate \ce{MAPbI3$\cdot$H2O}, where deep-lying Pb 5d levels are used as a reference for the valence band maximum (VBM) and conduction band minimum (CBM) of each phase.}
\end{figure}

Next, we derived thermodynamic transition levels $\varepsilon(q_1/q_2)$ between defects in different charge states $q_1$ and $q_2$. Figure~\ref{fig_trans} shows the possible transition levels together with the relative band alignment of \ce{MAPbI3\_H2O} and \ce{MAPbI3$\cdot$H2O} with respect to \ce{MAPbI3}. Water molecules inserted through the film surface extract electrons, resulting in the shift of the valence band maximum (VBM) towards lower values and the conduction band minimum (CBM) towards higher values, and thus the bandgap change from the pristine to the water-intercalated and to the monohydrated phase. It is apparent that in the case of \ce{MAPbI3} all of the vacancy defects have shallow transition levels, whereas in the case of hydrous compounds the vacancies exhibit deep trap behavior that can facilitate the recombination of charge carriers, resulting in the degradation of solar cell performance. Specifically, in the case of \ce{MAPbI3}, $V_I$ and $V_{\ce{MA}}$ are shallow donors and acceptors due to their transition levels $\varepsilon$(1+/0) and $\varepsilon$(0/1-) near the CBM and VBM, respectively. In the case of hydrous compounds, transition levels are located deep in bandgap region, although $V_{\ce{MA}}$ transition levels are not far away from the valence band. In particular, $V_{\ce{Pb}}$ has no transition level ($\varepsilon$(0/2-) inside the valence band) in \ce{MAPbI3}, but two deep transition levels $\varepsilon$(0/1-) and $\varepsilon$(1-/2-) are possible in the hydrous compounds. Similar features are found for $V_{\ce{MAI}}$ (two transition levels) and for $V_{\ce{PbI2}}$, which has two/four transition levels in the pristine/hydrous \ce{MAPbI3}.

%
\begin{figure}[!t]
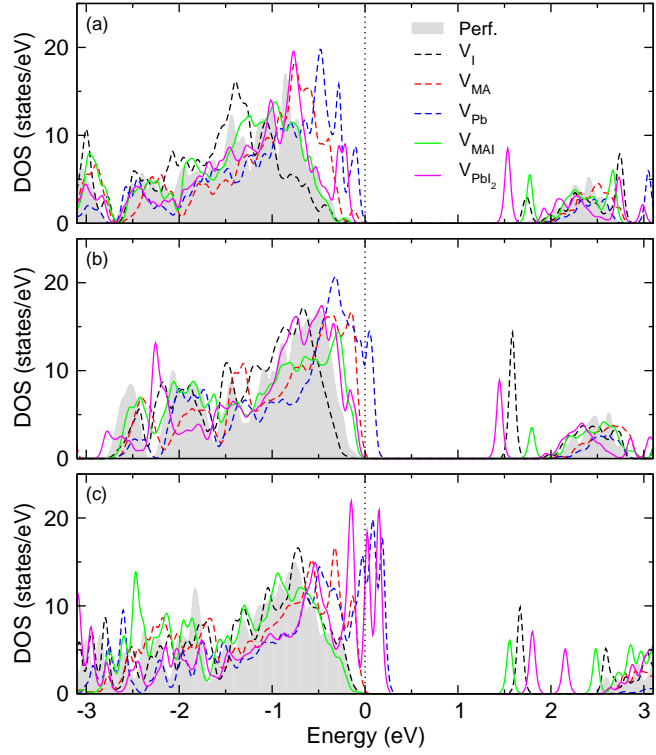

\begin{center}
\includegraphics[clip=true,scale=0.54]{fig3a.eps}
\includegraphics[clip=true,scale=0.54]{fig3b.eps}
\includegraphics[clip=true,scale=0.54]{fig3c.eps}
\end{center}
\caption{\label{fig_dos}Electronic density of states in perfect and vacancy-containing (a) \ce{MAPbI3}, (b) water-intercalated phase \ce{MAPbI3}\_\ce{H2O}, and (c) monohydrate phase \ce{MAPbI3$\cdot$H2O}. Using semi-core 5d levels of Pb far away from the defect as a reference, the VBM in the perfect phase is set to be zero, indicated by dotted vertical line. }
\end{figure}
The electronic density of states (DOS) in the pristine and vacancy-containing structures are shown in Figure~\ref{fig_dos}. Only the neutral states are considered. To make clear the role of each species, the atom-projected DOS (PDOS) are presented in Figure S4-S6. In all the compounds, the lower conduction band is from Pb 6p, while the upper valence band is from mostly I 5p states and minor part of Pb 6s ~\cite{yucj08,yucj10,yucj12}. In the case of \ce{MAPbI3},  $V_I$ formation causes a donor level near the conduction band and the positive potential results results in a downshift of the local valence band, while $V_{\ce{Pb}}$ and $V_{\ce{MA}}$ form acceptor states and their negative potential result in a local upshift of the valence band. The DOS characteristics of $V_{\ce{PbI2}}$ and $V_{\ce{MAI}}$ can be explained by combining the effects of individual point defects. Similar features are observed in the hydrous compounds. One distinction is that the n-type donor level created by $V_I$  is deeper in the band gap due to the presence of the water molecule. The electronic states of the water molecule overlap with I 5p at about $-2$ ($-3$) eV and with the MA states at about $-4$ ($-5$) eV in \ce{MAPbI3\_H2O} (\ce{MAPbI3$\cdot$H2O}), through the hydrogen bonding interaction between water and I atoms of \ce{PbI6} as well as MA moiety. When the vacancy defect $V_I$ or $V_{\ce{Pb}}$ is formed, similar interactions are observed.

\begin{figure*}[!t]
\scriptsize
\begin{center}
\begin{tabular}{ccc}
\includegraphics[clip=true,scale=0.17]{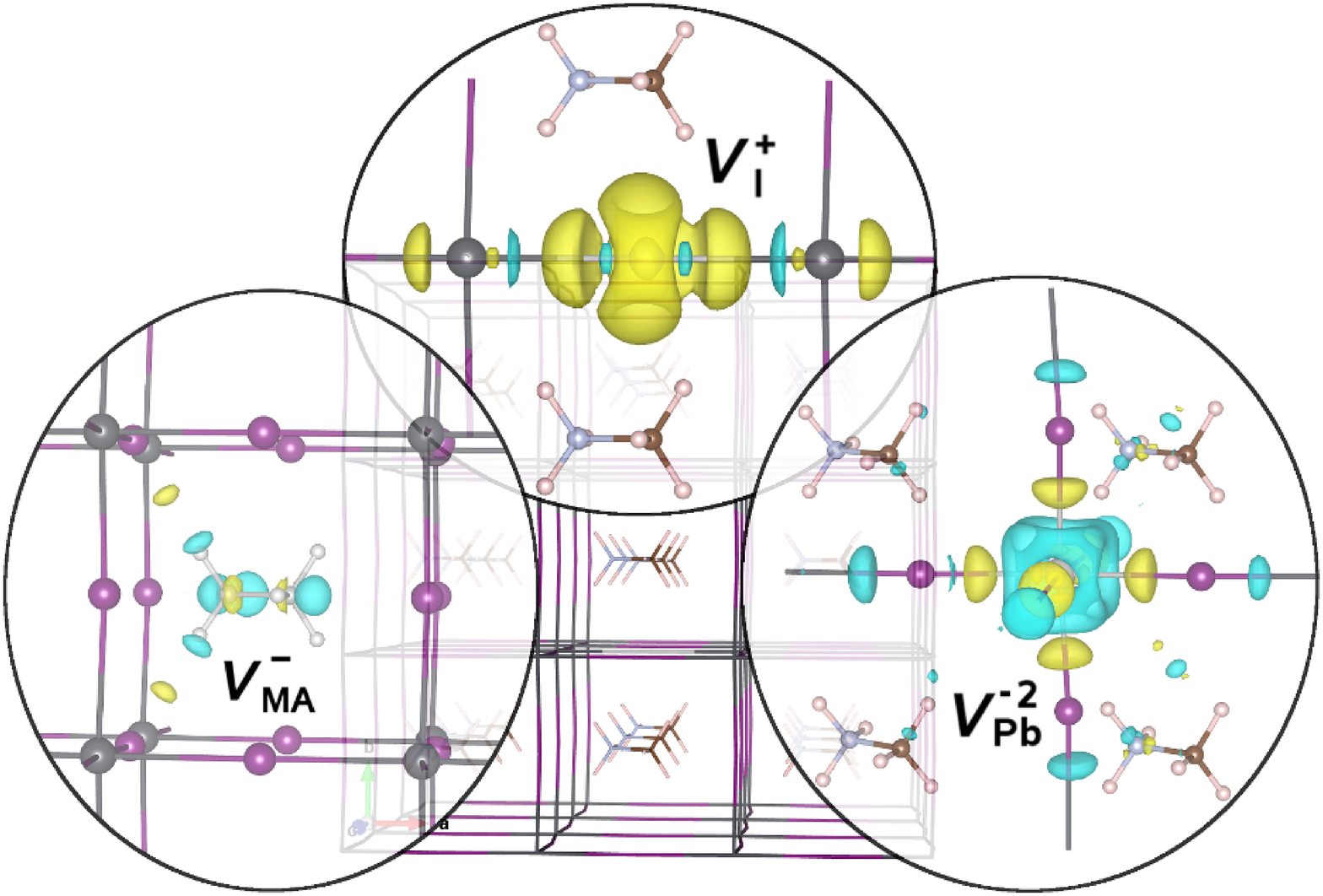} &
\includegraphics[clip=true,scale=0.17]{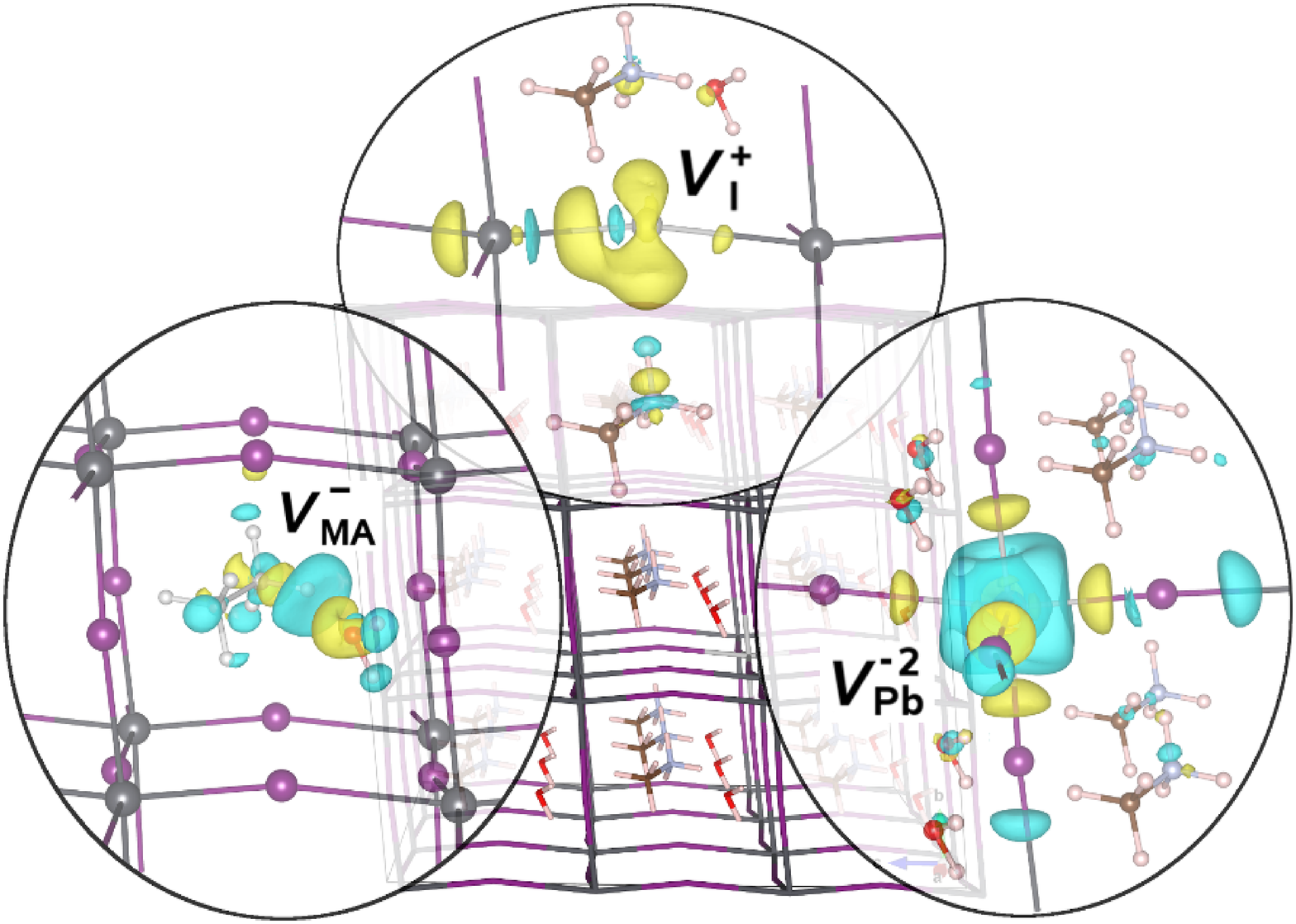} &
\includegraphics[clip=true,scale=0.17]{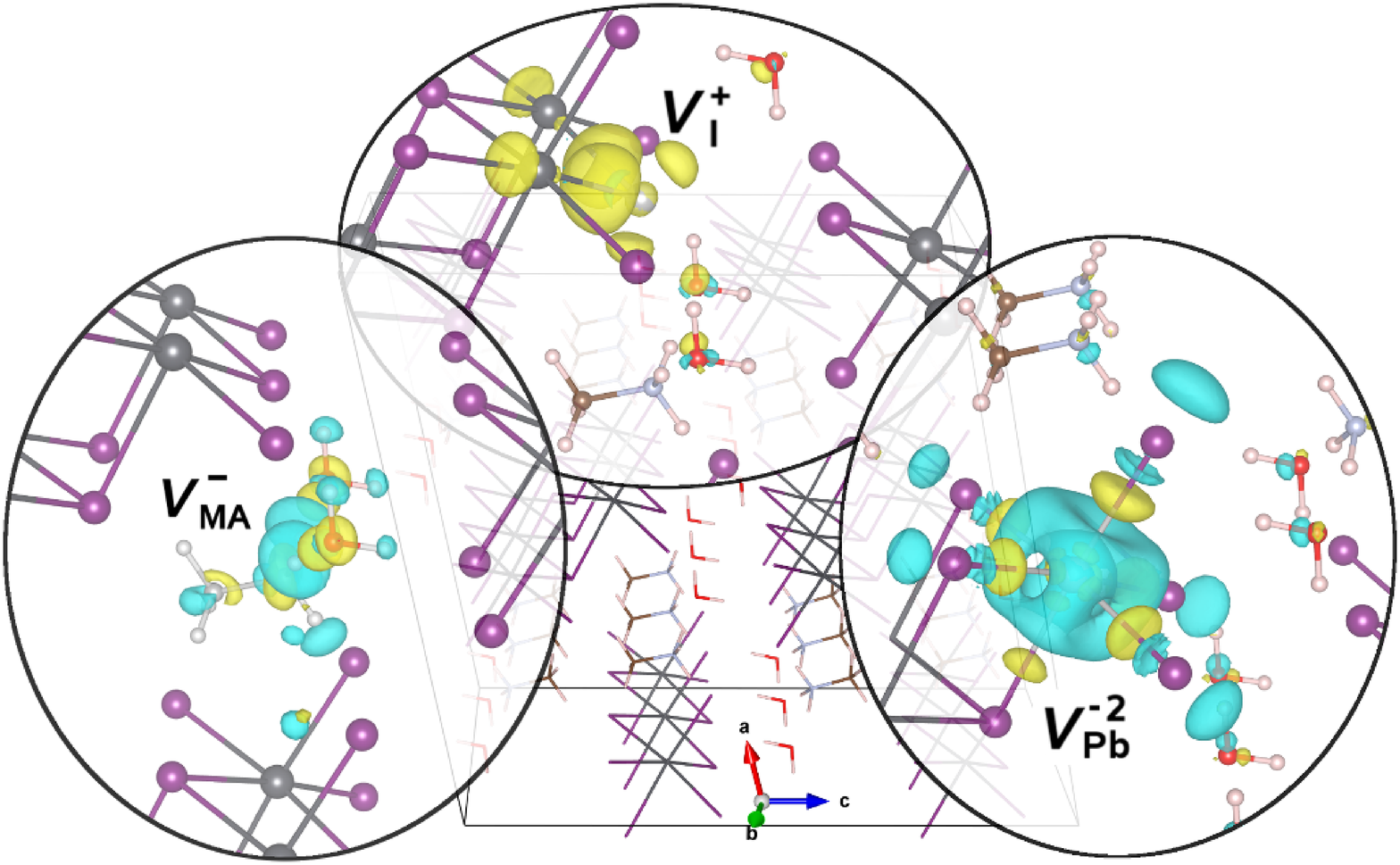} \\
(a) \ce{MAPbI3} & (b) \ce{MAPbI3\_H2O} & (c) \ce{MAPbI3$\cdot$H2O} 
\end{tabular}
\end{center}
\caption{\label{fig_den}Isosurface plot of electronic charge density difference upon formation of vacancy point defects $V_I^+$, $V_{\ce{MA}}^-$, $V_{\ce{Pb}}^{-2}$ in (a) \ce{MAPbI3}, (b) \ce{MAPbI3}\_\ce{H2O}, and (c) \ce{MAPbI3$\cdot$H2O}, at the value of 0.0025 $|e|$/\AA$^3$. Yellow (blue) color represents the charge depletion (accumulation).}
\end{figure*}

To study charge transfer during the formation of a vacancy defect, we plot the electron density difference $\Delta\rho=\rho_{\text{V}_\text{D}}-\rho_\text{perf}+\rho_\text{D}$ in Figure~\ref{fig_den}. In the case of \ce{MAPbI3},  charge is depleted around the $V^+_I$ defect, while charge is accumulated around $V^-_{\ce{MA}}$ and $V^{-2}_{\ce{Pb}}$, as expected. 
When the hydrous phases are formed, the extent of charge exchange is reduced because the water molecule can donate (for positive point defect) or accept (for negative ones) some electrons, indicating a stabilization by water  of the Schottky-type defects. 
In the absence of water, $V_I$, $V_{\ce{MA}}$ and $V_{\ce{Pb}}$ with the charge states of $+1$, $-1$ and $-2$ are the most stable over a large range of Fermi energy. In the case of hydrous compounds, however, these defects exchange the electrostatic stabilization by water opens up alternative charge states, leading to the creation of deep levels, which can enhance non-radiative processes and ultimately lead to the degradation of perovskite solar cells.

%

In summary, we have investigated defect processes in \ce{MAPbI3} and its hydrous phases \ce{MAPbI3\_H2O} and \ce{MAPbI3$\cdot$H2O} in order to reveal the effect of water on the performance and stability of iodide perovskites. The formation energies of $V_I$, $V_{\ce{MA}}$, $V_{\ce{Pb}}$, $V_{\ce{MAI}}$ and $V_{\ce{PbI2}}$ with various charge states, the binding energies of the complex defects ($V_{\ce{MAI}}$ and $V_{\ce{PbI2}}$), the density of states, and electronic charge density differences were presented to draw the following conclusions. 
The formation of $V_{\ce{PbI2}}$ from its individual vacancy point defects is spontaneous, and due to the greatly reduced kinetic barrier for \ce{I-} ion migration when hydrated, the concentration of $V_I$ should be reduced to prevent this formation, which can be realized by imposing  I-rich conditions. 
In the hydrous compounds, the formation of individual point defects $V_I$ and $V_{\ce{MA}}$ is more favorable than the formation of $V_{\ce{MAI}}$, and thus \ce{I2} or \ce{CH3NH2} or HI can be formed rather than MAI during the decomposition. Unlike in  bulk  \ce{MAPbI3}, all the vacancy defects create deep transition levels in the hydrous compounds arising from electrostatic interactions with water molecules. 
To overcome the negative effects of water on the performance and stability of halide perovskites, controlling the processing conditions such as 
the halide chemical potential during growth and annealing will be important, in additional to the physical encapsulation of devices.

\section*{Computational Methods}
The formation enthalpy of a point defect with a charge state $q$ is calculated using the grand canonical expression~\cite{Northrup,Freysoldt,Kumagai1},
\begin{equation}
\Delta H_f[D^q]\cong\{E[D^q]+E_\text{corr}[D^q]\}-E_\text{perf}-n_i\mu_i+qE_F \label{eq_Hf}
\end{equation}
where $E[D^q]$ and $E_\text{perf}$ are the total energies of the supercell including a defect $D$ and the perfect crystal supercell, and $n_i$ and $\mu_i$ are the number of removed (minus sign) or added (plus sign) $i$-type species and its chemical potential. $E_\text{corr}[D^q]$ is a correction to the error in the total energy of charged supercell that can be calculated by $E_\text{corr}=\alpha q^2/\varepsilon L$ in the monopole approximation, where $\alpha$ is the Madelung constant, $\varepsilon$ the static dielectric constant, and $L$ the lattice constant, respectively~\cite{monopole,Freysoldt}. Using the density functional perturbation theory, we computed the isotropic static dielectric constants as 23.55, 25.88, and 16.30 for \ce{MAPbI3}, \ce{MAPbI3}\_\ce{H2O}, and \ce{MAPbI3$\cdot$H2O}. $E_F$ is the Fermi energy expressed as referenced to the valence band: $E_F=\epsilon_\text{VBM}+\Delta\epsilon_F+\Delta V$, where $\epsilon_\text{VBM}$ is the energy level of the VBM, $\Delta\epsilon_F$ is the Fermi level with respect to the VBM and $\Delta V$ is the potential alignment. We ignored the band-filling correction for shallow defects due to its negligible value with large supercell size in this work. Self-interaction and spin-orbit coupling were not included, which are not expected to affect the comparison between the defect physics of pristine and hydrated \ce{MAPbI3} compounds and thereby draw the conclusion about the decomposition mechanism upon hydration, but they will be important for quantitative defect spectroscopy~\cite{MHDu}.

The chemical potentials depend on the growth conditions, which can fall between  I-rich or I-poor conditions. The I-rich condition corresponds to the iodine precursor in orthorhombic solid form (space group $Cmca$), and thus the upper limit of the iodine chemical potential is that $\mu_I^\text{rich}=E_{\ce{I}(\text{orth})}$. The synthesis equations constrain the chemical potentials as follows,
\begin{gather}
\mu_{\ce{MAI}}+\mu_{\ce{PbI2}}=\mu_{\ce{MAPbI3}} \label{eq_mai} \\
\mu_{\ce{Pb}}+2\mu_{\ce{I}}=\mu_{\ce{PbI2}} \label{eq_pb} \\
\mu_{\ce{MA}}+\mu_{\ce{I}}=\mu_{\ce{MAI}} \label{eq_ma}
\end{gather}
Eqs.~(\ref{eq_mai}) and (\ref{eq_pb}) correspond to the real synthetic reactions, but Eq.~(\ref{eq_ma}) acts only as a theoretical reference. 
From Eq.~(\ref{eq_pb}), we identified the iodine poor conditions, $\mu_I^{\text{poor}}\approx1/2(E_{\ce{PbI2}}-\mu_{\text{Pb}}^{\text{rich}})$, where $\mu_{\text{Pb}}^{\text{rich}}=E_{\ce{Pb}(\text{fcc})}$ 
referred to the bulk Pb in fcc phase and $E_{\ce{PbI2}}$ is referred to the bulk \ce{PbI2} in rhombohedral phase (space group $P$\={3}$m1$)~\cite{yucj12}, and Pb-poor condition, $\mu_{\text{Pb}}^{\text{poor}} \approx E_{\ce{PbI2}}-2\mu_{\ce{I}}^{\text{rich}}$. 
Then, using Eqs.~(\ref{eq_mai}) and~(\ref{eq_ma}), the MA-poor and -rich conditions are established like $\mu_{\ce{MA}}^{\text{poor}} \approx E_{\ce{MAPbI3}}-E_{\ce{PbI2}}-\mu_{\ce{I}}^{\text{rich}}$ and $\mu_{\ce{MA}}^\text{rich}\approx E_{\ce{MAPbI3}}-E_{\ce{PbI2}}-\mu_{\ce{I}}^\text{poor}$. 
For the vacancy pair defects, the chemical potentials were not affected by the iodine chemical potential; $\mu_{\ce{MAI}} \approx E_{\ce{MAPbI3}}-E_{\ce{PbI2}}$ and $\mu_{\ce{PbI2}} \approx E_{\ce{PbI2}}$.

Pseudo-cubic unit cells were adopted for \ce{MAPbI3}, and $(3\times3\times3)$ and $(2\times3\times2)$ supercells were used for vacancy-containing \ce{MAPbI3}, \ce{MAPbI3\_H2O} and \ce{MAPbI3$\cdot$H2O}, respectively. The DFT total energies were calculated using the {\small Quantum ESPRESSO} code~\cite{QE} with the ultrasoft pseudopotentials provided in the code and the Perdew-Burke-Ernzerhof (PBE) exchange-correlation functional~\cite{pbe} added by the van der Waals (vdW) energy in the flavor of vdW-DF-OB86~\cite{vdWDFob86}. Scalar-relativistic effects are included.
A plane-wave cutoff energy of 40 Ry and a $\Gamma$ point for structural relaxation of vacancy-containing supercells, while $2\times2\times2$ special $k$-points for DOS calculations, were used for all the configurations. All the atomic positions of each configuration were relaxed until the forces on atoms converge to $5\times 10^{-5}$ Ry/Bohr. 

\section*{Appendix A. Supplementary data}
Supplementary data related to this article can be found at URL.

\section*{\label{note}Notes}
The authors declare no competing financial interest.

\bibliographystyle{elsarticle-num-names}
\bibliography{Reference}

\end{document}